\documentclass[10pt,journal]{IEEEtran}

\usepackage{setspace}
\usepackage{cite,graphicx,amsmath,amssymb}
\usepackage{fancyhdr}

\usepackage{mdwmath}
\usepackage{mdwtab}
\usepackage{balance}
\usepackage{xcolor}
\usepackage{bm}
\usepackage{amsthm}
\usepackage{threeparttable}
\usepackage{algorithm}
\usepackage{algorithmic}
\usepackage{multirow}
\usepackage{flafter}
\usepackage{makecell}
\usepackage{diagbox}
\usepackage{xcolor}
\usepackage{epstopdf}
\usepackage{graphicx}
\usepackage[numbers]{natbib}
\usepackage{setspace}
\usepackage{subcaption}

\usepackage{algorithm}
\usepackage{algorithmic}

\begin{document}
\title{Intelligent Travel Activity Monitoring: Generalized Distributed Acoustic Sensing Approaches}

\author{\normalsize {
Ruikang~Zhong,~\IEEEmembership{\normalsize Member,~IEEE,}
Chia-Yen~Chiang,~\IEEEmembership{\normalsize Graduated Student Member,~IEEE,}\\
Mona~Jaber,~\IEEEmembership{\normalsize Senior~Member,~IEEE,}
Rupert De Wilde, Peter Hayward.
}

\thanks{Ruikang~Zhong, Chia-Yen~Chiang, and Mona Jabor are with the School of Electronic Engineering and Computer Science, Queen Mary University of London, London E1 4NS, U.K. (e-mail: r.zhong@qmul.ac.uk; c.chiang@qmul.ac.uk; m.jaber@qmul.ac.uk;).

Rupert De Wilde is with Project Delivery Lead at Sensonic Ltd., GU14 7NA, Farnborough, U.K. (e-mail: rupertdewilde@hotmail.com)

Peter Hayward is with T2 Sensing Ltd., 4 Richmond Road, Grays, RM17 6DW, U.K. (e-mail: peter.hayward@t2sensing.com)
}
}


\date{\today}
 \maketitle

\begin{abstract}
Obtaining data on active travel activities such as walking, jogging, and cycling is important for refining sustainable transportation systems (STS). Effectively monitoring these activities not only requires sensing solutions to have a joint feature of being accurate, economical, and privacy-preserving, but also enough generalizability to adapt to different climate environments and deployment conditions. In order to provide a generalized sensing solution, a deep learning (DL)-enhanced distributed acoustic sensing (DAS) system for monitoring active travel activities is proposed. By leveraging the ambient vibrations captured by DAS, this scheme infers motion patterns without relying on image-based or wearable devices, thereby addressing privacy concerns. We conduct real-world experiments in two geographically distinct locations and collect comprehensive datasets to evaluate the performance of the proposed system. To address the generalization challenges posed by heterogeneous deployment environments, we propose two solutions according to network availability: 1) an Internet-of-Things (IoT) scheme based on federated learning (FL) is proposed, and it enables geographically different DAS nodes to be trained collaboratively to improve generalizability; 2) an off-line initialization approach enabled by meta-learning is proposed to develop high-generality initialization for DL models and to enable rapid model fine-tuning with limited data samples, facilitating generalization at newly established or isolated DAS nodes. Experimental results of the walking and cycling classification problem demonstrate the performance and generalizability of the proposed DL-enhanced DAS system, paving the way for practical, large-scale DAS monitoring of active travel.

\end{abstract}

\begin{IEEEkeywords}
Active travel, deep learning, distributed acoustic sensor, federated learning, meta learning.
\end{IEEEkeywords}

\section{Introduction}

Active travel modes, such as walking, jogging, cycling, and skateboarding, play a crucial role in sustainable transportation systems (STS), which not only reduce the congestion caused by vehicle transportation but also benefit public health \cite{10700813, brand2021climate}. Accurately gathering the information of active travel activities is essential for transportation management, urban planning, and the development of a smart city \cite{alattar2021public}. The utilization of Internet-of-Things (IoT) devices and sensing technology to collect travel information has emerged as a promising solution, demonstrating superior efficiency and accuracy in comparison to conventional questionnaires or manual statistics. However, the prevalent image-based methods or invasive devices often compromise individual privacy \cite{9716851}, limiting their application scenarios and public acceptance in practical applications. Consequently, there is an increasing demand for efficient and privacy-preserving sensing technologies for travel activity detection in transportation systems.

Fortunately, the distributed acoustic sensing (DAS) system integrated with advanced artificial intelligence (AI) techniques offers a promising solution for active travel monitoring. Overall, in this scheme, the DAS sensors can collect acoustic information caused by the activity, while the deep learning~(DL) algorithm is responsible for the analysis of collected signals~\cite{10843348}. Specifically, the DAS technology re-utilizes existing fiber-optic cables to detect human activities without recording personal biometric data (e.g., appearance), thus inherently addressing privacy concerns \cite{corera2023long}. Furthermore, the DAS system has a coverage area that encompasses tens or even hundreds of kilometers, and in the time domain, its monitoring is long-term and uninterrupted surveillance \cite{9896929}. Once the DAS sensors are integrated with well-trained DL models, DAS enables fine-grained interpretation of complex activity patterns, demonstrating significant superiority in terms of coverage, accuracy, privacy, and efficiency for monitoring active travel motions~\cite{8891715}.

Despite the promise of AI-enhanced DAS systems, several practical challenges remain unresolved due to the variability of deployment environments at different DAS nodes in large-scale DAS implementation. Since optical fibers are buried underground, each DAS implementation is influenced not only by the status of the fiber itself but also by environmental factors, including but not limited to the road conditions, temperature, humidity, and soil hardness \cite{wu2022improved}. Such heterogeneity undermines the generalization capability of DL models that are well-fitted to any specific environment. Although it is a feasible solution to perform independent local training for each DL model employed by different DAS nodes, the hardware expense, time cost, and computational complexity required for collecting the local dataset and conducting complete training are enormous and unaffordable. Therefore, addressing the challenge of generalizability in AI-enhanced DAS systems is a prerequisite for the widespread deployment of the system. To address the generalization issue in DAS practice, in this paper, we propose two approaches based on federated learning (FL) and meta-learning, which fit different scenarios depending on the network availability. By enabling models to collaboratively learn from decentralized data or training a generalized initialization that can quickly adapt to new environments, the proposed schemes ensure the accuracy of active travel activity classification across diverse DAS deployments at the same accurate level of local data training.

\begin{figure*}[t!]
\centering
\includegraphics[width=0.8\textwidth]{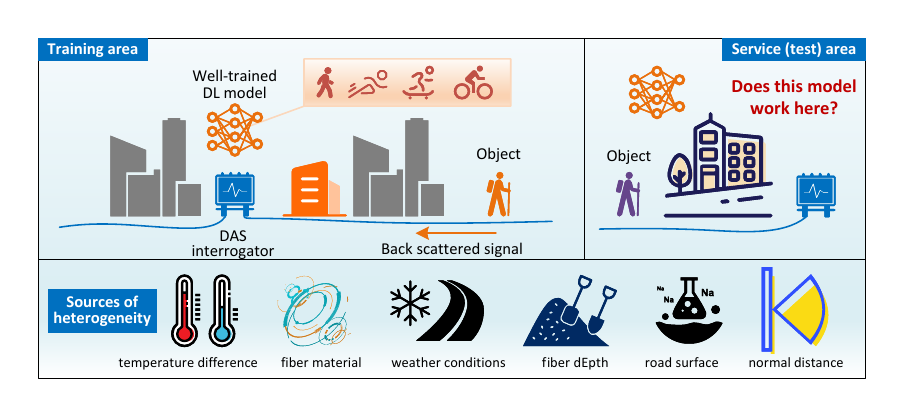}
\caption{Model and working principle of the DAS system.}
\label{Fig.1}
\end{figure*}

\vspace{-0.3cm}
\subsection{Related Research}
\subsubsection{DAS enabled ITS}

The DAS system boasts extensive coverage and economic advantages, making it a successful solution for a variety of applications, such as pipeline monitoring~\cite{10075395}, earthquake detection \cite{lior2023magnitude}, and perimeter detection~\cite{9953529}. In the realm of transportation, a series of studies have been formulated with the objective of employing DAS systems to monitor the operation of vehicles \cite{10368099} and trains~\cite{wiesmeyr2020real}. Vehicle classification is a classic application of DAS in transportation systems. A vehicle detection and classification algorithm based on signal processing and feature extraction was proposed in~\cite{8943448}, with an accuracy of 80\% and 70\%, respectively. In \cite{8453776}, a wavelet threshold algorithm was proposed to support DAS systems in estimating the speed of multiple vehicles. Furthermore, the authors of~\cite{9564517} proposed an image processing scheme to estimate the traffic flow and average speed of vehicles using image representations generated from segments of DAS signals.

The emergence of machine learning technologies further assisted the DAS system, and integrating DAS and data-driven machine learning methods exhibits potential to develop AI-enhanced traffic perception systems~\cite{10644017}.
Deep learning-based algorithms such as convolutional neural networks~\cite{10078155}, recurrent neural networks, and transformers~\cite{10556648} are helpful to extract spatiotemporal features from DAS signals to achieve tasks such as vehicle type discrimination, pedestrian behavior recognition, and road anomaly detection~\cite{shao2025artificial}. A U-Net image segmentation model was proposed to detect low-speed quasi-static vehicles \cite{10904069}. In our previous work \cite{10843348}, a sparse residual network (SR-Net) model and an Alex-SR model were proposed for the vehicle classification and occupancy detection problem, respectively. This approach solves the problem that the low strain of quasi-static vehicles is easily masked by environmental noise and DAS fading noise. However, it is worth noting that the current applications of DAS focus on heavy equipment, while DAS detection is more challenging for active travelers who have significantly lower weight.

\subsubsection{Generalization of Models}
The effects of ambient temperature, road material, deployment depth, and fiber usage time on the received signal-to-noise ratio of the DAS system are revealed in \cite{10368099}. The authors found that the above factors can have an impact of more than 3dB on the received signal-to-noise ratio through vehicle experiments, which supports the necessity of improving the generalization capability of models. The authors of \cite{wu2022improved} proposed an unsupervised spiking neuron network to improve the generalization capability of analyzing DAS signals. Since there are limited existing studies on improving the generalizability of DAS systems, we try to find solutions in the field of computer science. Although FL was originally proposed to improve the training efficiency of the model, some studies have indicated that it has a positive effect on improving the generalizability of the model \cite{10571602}. The statements in \cite{mora2024enhancing} also pointed out that FL can train a global model with good generalization ability when there is bias among local data. The authors of \cite{10011632} proposed using FL in distributed IoT devices that independently collect their local data, and the model is capable of learning universal features, which inspired us to invoke FL among DAS nodes to improve the generalizability.

Conversely, meta-learning has been demonstrated to facilitate the training of generalized models across a variety of tasks and enable rapid adaptation to novel ones \cite{9428530}. There are three common types of meta-learning, namely metric-based, model-based, and optimization-based meta-learning \cite{10413635}. Given the heterogeneity of tasks in DAS systems, optimization-based primitive learning is particularly well-suited, as it can be compatible with a range of models, as opposed to being confined to a particular architecture and task. Model-Agnostic Meta-Learning (MAML) is a classic optimization based meta-learning algorithm that enables the model to quickly adapt to new tasks after a few gradient updates by finding shared initialization parameters on multiple tasks \cite{pmlr-v70-finn17a}. However, MAML has a high computational complexity. Fortunately, Reptile \cite{nichol2018first} is a simplified method of MAML, which does not need to calculate the second-order gradient, but performs several steps of ordinary gradient descent on multiple tasks, which is more suitable for DAS nodes with limited computing resources.

\vspace{-0.3cm}
\subsection{Our Contribution}

As demonstrated in the preceding review of current research, there are still several challenges with regard to the monitoring of intelligent travel modes. Firstly, the existing research on monitoring active travel modes using IoT devices and sensing technology is insufficient. Furthermore, despite the numerous successful applications of DAS, and the proposed utilization of DAS providing sensing functions within traffic systems as outlined in several papers, there remains a paucity of research focusing on DAS in active transportation, particularly in the realm of experimental studies. Finally, the issue of intelligent agent performance degradation resulting from heterogeneity in the physical and geographical environment remains unresolved. This is particularly pertinent for systems such as DAS with a vast coverage area, where the absence of the generalized model can result in a significant deterioration in performance during practical implementation.

In order to resolve the above mentioned technical challenges, we propose an AI-enhanced DAS system to monitor the activities of active travelers. Two different solutions are proposed to address the generalization challenges in employing DL approaches in the DAS monitoring system. In summary, this paper makes the following original contributions

\begin{itemize}
    \item \textbf{DAS system:} We propose an active travel activity classification framework that leverages DAS and DL technologies to identify active travelers' activities in an efficient and privacy-preserving manner. The proposed method utilizes ambient vibration signals captured by DAS to infer human motion patterns.

    \item \textbf{Experiments:} To validate the effectiveness of the DAS system, we conduct physical experiments at two geographically distinct locations. We collected three datasets of DAS signals under various pedestrian activity scenarios, providing a foundation for the development of generalized DL models.

    \item \textbf{FL-based IoT solution:} We design a FL-based generalization scheme tailored for DAS nodes connected by IoT networks. This framework enables models deployed in different regions to be trained collaboratively by sharing model parameters, thereby accelerating convergence and improving cross-region generalizability.

    \item \textbf{Meta learning based initialization:} To address generalization issues in scenarios with restricted connectivity or when initializing new DAS infrastructure, we invoke meta-learning and few-shot learning techniques. These methods enable the development of generalized models capable of rapidly adapting to local conditions with only a few local data samples.
\end{itemize}

\vspace{-0.2cm}
\section{DAS Enabled Active Traveling Detection}\label{Section2}

The main function of the proposed AI-enhanced DAS system is to perceive the movement patterns of active travelers. Without loss of generality, in this paper, we consider classifying the two most common active travel modes, walking and cycling. As shown in Fig. \ref{Fig.1}, two sets of DAS systems are deployed in different cities, and each of them consists of three components: a DAS interrogator, underground fiber, and a DL model. Specifically, the DAS interrogator is responsible for transmitting pulse signals into the optical fiber and receiving backscattering signals. By observing the phase shift of the received reflected signal, the DL model can extract key features and accordingly detect the type of movement causing this disturbance. For a concise presentation, we denote an independent set of a DAS interrogator,  fibers linked to the interrogator, and corresponding DL models as a DAS node. In the rest of this section, we will introduce the theory of DAS technology, the DL model, and the challenges of generalizability.

\vspace{-0.2cm}
\subsection{DAS Basics}

Connecting a DAS interrogator transforms dark optical fibers into an array of virtual sensors, enabling continuous monitoring of acoustic events over the periphery of the fiber. Since optical fibers are often laid along roadsides near bicycle lanes and pedestrian paths, it is a natural advantage to use them to monitor active travel. The interrogator sends light pulses $E = E_0 e ^ { - j \omega t }$ through the fiber, where $E_0$ represents the amplitude of the transmitted signal and $\omega$ represents the frequency of the signal, respectively. The light pulses through the fiber interact with the medium and generate a Rayleigh backscattering signal $E_b = {E_0} \beta e ^ { - j \omega t}$ at discrete segments, which are known as fiber bins, where the number of bin is denoted by $b$. Each bin reflects part of the signal with a specific phase and amplitude, forming the basis of signal analysis. $E_b$ and $E$ have opposite propagation directions, and $E_b$ has a certain degree of amplitude attenuation, where $\beta$ is the amplitude attenuation coefficient, which depends on multiple factors such as fiber length and Rayleigh backscattering coefficient.

Without external disturbances, the phase of the backscattered signal remains stable, which is only affected by noise. Once an acoustic event happens (e.g., a pedestrian step on the ground), the vibration of the ground and fiber results in a phase shift on the back scattering signal, which can be expressed as  $E_b = {E_0} \beta e ^ { - j \omega t+\vartriangle \varphi_{b}}$. The principle is that vibrations caused by acoustic events strain the fiber, altering both its length and refractive index, and thus the phase shift of the backscattered signal is changed as a result.

The DAS interrogator detects these phase variations by comparing the interference intensity over time. This allows it to track dynamic phase changes and reconstruct a time-series signal for each fiber bin. The received DAS signal at bin $b$ is given by
\begin{align}
\mathbf{x}_{b,t} = [\vartriangle \varphi_{b,t}, \vartriangle \varphi_{b,t+1}, ...\vartriangle \varphi_{b, t+W}],\label{eq1}
\end{align}
where $W$ is the length of the window in the time domain. In practice, combining signals from multiple bins is more likely to help achieve more accurate decisions. Thus, it is also possible to sample in multiple bins to obtain a two-dimensional matrix signal
\begin{equation}
\mathbf{X}_{b,t} =
\begin{bmatrix}
\vartriangle \varphi_{b,t} & \vartriangle \varphi_{b,t+1} & \cdots & \vartriangle \varphi_{b,t+W} \\
\vartriangle \varphi_{b+1,t} & \vartriangle \varphi_{b+1,t+1} & \cdots & \vartriangle \varphi_{b+1,t+W} \\
\vdots & \vdots & \ddots & \vdots \\
\vartriangle \varphi_{b+H,t} & \vartriangle \varphi_{b+H,t+1} & \cdots & \vartriangle \varphi_{b+H,t+W}\label{eq2}
\end{bmatrix}
\end{equation}
where $H$ is the window size in the spatial domain. Taking the variations of $\mathbf{X}_{b,t}$ as a clue, the DL agent is able to infer the movement of the traveler,

\vspace{-0.2cm}
\subsection{Independent DL Model}
The basic DL models work in an independent manner to classify the type of active travel, which is a binary classification problem. Given the input matrix $\mathbf{X} \in \mathbb{C}^{H \times W}$ constructed from DAS signals, the goal is to estimate the corresponding label $y \in \{0,1\}$, which denotes the label of walking and cycling, respectively. Denoting the estimated label by $\hat{y} = f_\omega(\mathbf{X})$, where $f_\omega(\cdot)$ is a trainable model parameterized by $\omega$. The training objective of the DL model aims to find out a set of trainable parameters to maximize the accuracy of the classification over the dataset $\mathcal{D} = \{(\mathbf{X}^{n}, y^{n})\}_{n=1}^N$, which can be formulated as
\begin{subequations}
\begin{align}\label{OPPY}
\mathcal{P}1: &\max_\omega \ \frac{1}{N} \sum_{n=1}^{N} \mathbb{I}(f_\omega(\mathbf{X}^{n}) = y^{n}) \\
\textrm{s.t.} \ \
& y^n, \hat{y}^n \in\{0,1\}, \forall n, \label{OPP1}
\end{align}
\end{subequations}
where $\mathbb{I}(\cdot)$ is the indicator function, returning 1 if the prediction is correct and 0 otherwise, and constraint \eqref{OPP1} is the constraint of the legal labels.

\begin{figure}[t!]
\centering
\includegraphics[width=0.5\textwidth]{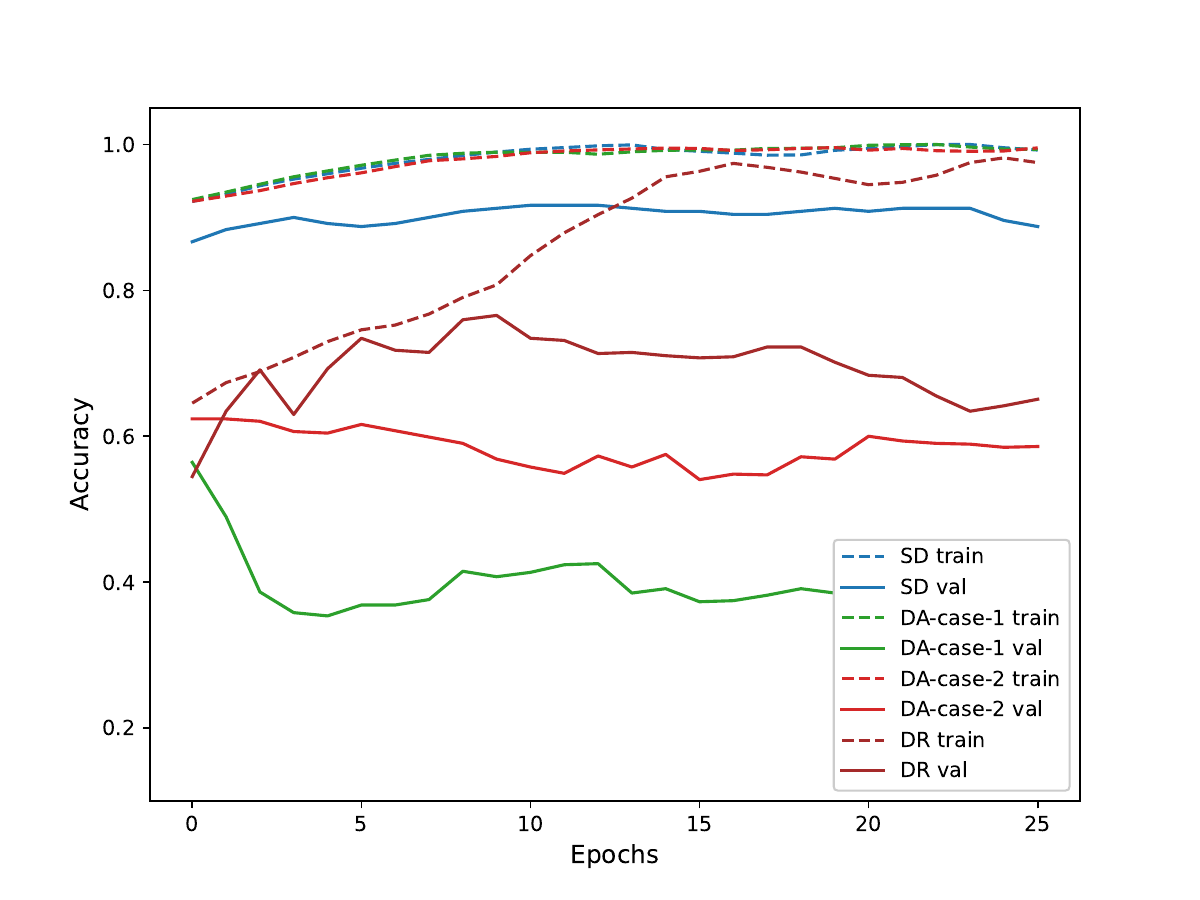}
\caption{Performance of training the SR-Net with data from different DAS nodes.}
\label{Fig.2}
\end{figure}

To address the formulated classification task, we employ \textbf{sparse residual network (SR-Net)}, a lightweight and efficient DL model designed for the DAS signal. The structure of SR-Net was proposed in our previous work \cite{10843348}, and the primary purpose of it is for the vehicle classification problem. It is re-employed in this work as it also demonstrated outstanding performance for the active travel scenarios as shown in Fig.~\ref{Fig.2}. The input to SR-Net is $\mathbf{X}_{b,t}$ that formed by stacked DAS signals from consecutive virtual sensors.

SR-Net is inspired by ResNet~\cite{He_2016_CVPR}, which introduces residual connections to mitigate the degradation problem in DL models. However, the traditional ResNets can have 50-200 layers, such a huge model is unnecessary for the DAS signal and over-complicated for the edge devices at the DAS nodes. To address the complexity problem, we introduce SR blocks, which better fit the network having fewer layers while retaining expressive power. Each sparse block applies multiple convolutional layers with non-linear activations, which is given by
\begin{equation}
f(\mathbf{X}_{i+4}) = f_{\omega,\sigma,{i+4}} (f_{\omega,\sigma,{i+3}} (f_{\omega,\sigma,{i+2}} (f_{\omega,\sigma,{i+1}} (\mathbf{X}_i)))) + \mathbf{X}_i
\end{equation}
where $\sigma$ denotes the activation function (e.g., ReLU) and $i$ represents the layer number. The overall architecture of SR-Net consists of only 26 trainable layers. Compared with conventional ResNet models, this design significantly reduces computational complexity while maintaining robust performance for processing DAS signals.

\subsection{Challenges of Model Generalization Across DAS Nodes}

While SR-Net demonstrates outstanding performance for the classification problem within a local dataset, its generalization capability across different DAS nodes remains limited. In practical deployments, DAS signals captured from different geographic locations or fiber segments often exhibit significant variation for the same acoustic events. The reason is that DAS nodes have to be located in different areas to ensure coverage, and the distance between them is large enough to cause geographical heterogeneity. Therefore,  the following factors cause significant differences in the phase shift of the received DAS signal at DAS nodes.

\subsubsection{Environmental Conditions}
Since the principle of the DAS system relies on the deformation of the optical fiber and the soil around it, the road material and soil quality in different regions will directly affect the DAS signal. Furthermore, climate factors such as temperature, humidity, and weather conditions can have a substantial impact on the propagation of ground vibrations and the optical fiber’s sensitivity to acoustic events.

\subsubsection{Fiber Material and Condition}
The quality of optical fibers may vary in terms of core composition and manufacturing precision. These physical differences directly affect the attenuation factor $\alpha$ in the Rayleigh backscattering. In addition, the age and condition of the fiber can also cause different degrees of distortion and attenuation during the transmission of the DAS signal.

\subsubsection{Fiber Deployment Conditions}
The deployment setup of the optical fiber, such as burial depth and the distance to the road surface, also plays a crucial role in the DAS practice. Due to the sensitivity of vibration to distance, changes in burial depth or horizontal offset from the subject and fiber can result in significantly different signal amplitudes and temporal profiles. As a consequence, models trained on one deployment configuration may not be able to generalize to another case.

Due to above mentioned factors, a model trained on data from one DAS node may fail to reach the same accuracy when directly applied to another DAS node. The above problems are challenging to solve by analytical solutions such as scaling, filtering, etc., since these practical issues are not analytical. Fig. \ref{Fig.2} provides numerical evidence to prove the existence of this challenge. We plot model accuracy results for three cases, namely
\begin{itemize}
\item \textbf{Same Data Set (SD):} The SR-Net model is trained and tested using training and test datasets from the same area.
\item \textbf{Different Areas (DA):} The training and test datasets used by the SR-Net model are sampled from two regions with significant geographical differences.
\item \textbf{Different Road Sections (DR):} The training and testing datasets for the SR-Net model are sampled from different road sections in the same region.
\end{itemize}

It can be observed in Fig. \ref{Fig.2} that in terms of the SD case, the accuracy of the model exceeds $90\%$, which is an encouraging result. However, the performance of the model in the DA and DR cases dropped significantly, to $40\%$ and $60\%$, respectively. It is worth noting that this is a binary classification problem, and a $40\%$ accuracy rate indicates that the model is not only rambling, but even has a negative effect. This lack of generalizability can significantly increase the training cost of DL models for DAS systems as every DAS node need to train its own model. Therefore, it is necessary to further explore training strategies or calibration techniques to enhance the model’s generalizability and adaptability across heterogeneous DAS environments.

\subsection{Problem formulation}

In order to solve the above-mentioned generalizability problem of the DL model in the DAS system, we extend the classification problem of an independent DAS node described in eq. \eqref{OPPY} to a multi-nodes case. We assume there are a collection of $M$ DAS nodes distributed at different areas, each of them collects its local dataset $\mathcal{D}^{m} = {(\mathbf{X}^{n,m}, y^{n,m})}_{n=1}^{N_m}$, where $m \in {1, 2, ..., M}$ and $N_m$ represents the number of samples in dataset $\mathcal{D}^{m}$. The goal is to train a unified model $f_\omega$ to maximizes the average prediction accuracy across all DAS nodes, despite the variability in signal characteristics introduced by environmental conditions, fiber materials, and deployment differences. The objective can be defined as
\begin{subequations}
\begin{align}\label{OPP}
\mathcal{P}1: &\max_\omega \frac{1}{\sum_{m=1}^M N_m} \sum_{m=1}^M \sum_{n=1}^{N_m} \mathbb{I}(f_\omega(\mathbf{X}^{n,m}) = y^{n,m}) \\
\textrm{s.t.} \ \
& y^n, \hat{y}^n \in\{0,1\}, \forall n, \label{OPPZ1}\\
& N^m >= 5,  \forall m, \label{OPP1}
\end{align}
\end{subequations}
where \eqref{OPP1} is the constraint of data samples, ensuring each node has at least 5 data samples in the local setup. This formulation emphasizes the demand for a generalized model that can perform reliably across heterogeneous DAS conditions, rather than optimizing solely for an individual case.

\section{Experiment and Data Collection}

To obtain the data set for the research of the generalization capability of the DAS system, we conducted experiments on two independent DAS nodes at \textbf{Redfields Lane} in Church Crookham and \textbf{Cellarhead Junction}. These two nodes differ significantly in terms of environmental conditions, fiber materials, and deployment configurations, providing a representative testbed for cross-node adaptation.


\subsection{Experiment Setup}

The Helios interrogator was employed to gather the DAS signal following the principle described in Section \ref{Section2}. We first conducted an experiment in Redfields Lane, which is located in Crondall in southern England. Two subjects participated in the data collection of both experiments, and they had different weights and exercise characteristics. The deployment of fiber in Redfields Lane is below 0.5 meters deep by the edge of the road as shown in Fig. 3(a). The total length of the optical fiber is about 2.4 kilometers, and the DAS interrogator is deployed at one end of the optical fiber. During the experiment, the subjects moved back and forth along the blue highlighted Redfields Lane in Fig. 3(a), using their mobile phones to obtain GPS positioning data.

\begin{figure}[htb]
    \centering
    \begin{subfigure}[b]{0.45\textwidth}
        \centering
        \includegraphics[width=0.9\linewidth]{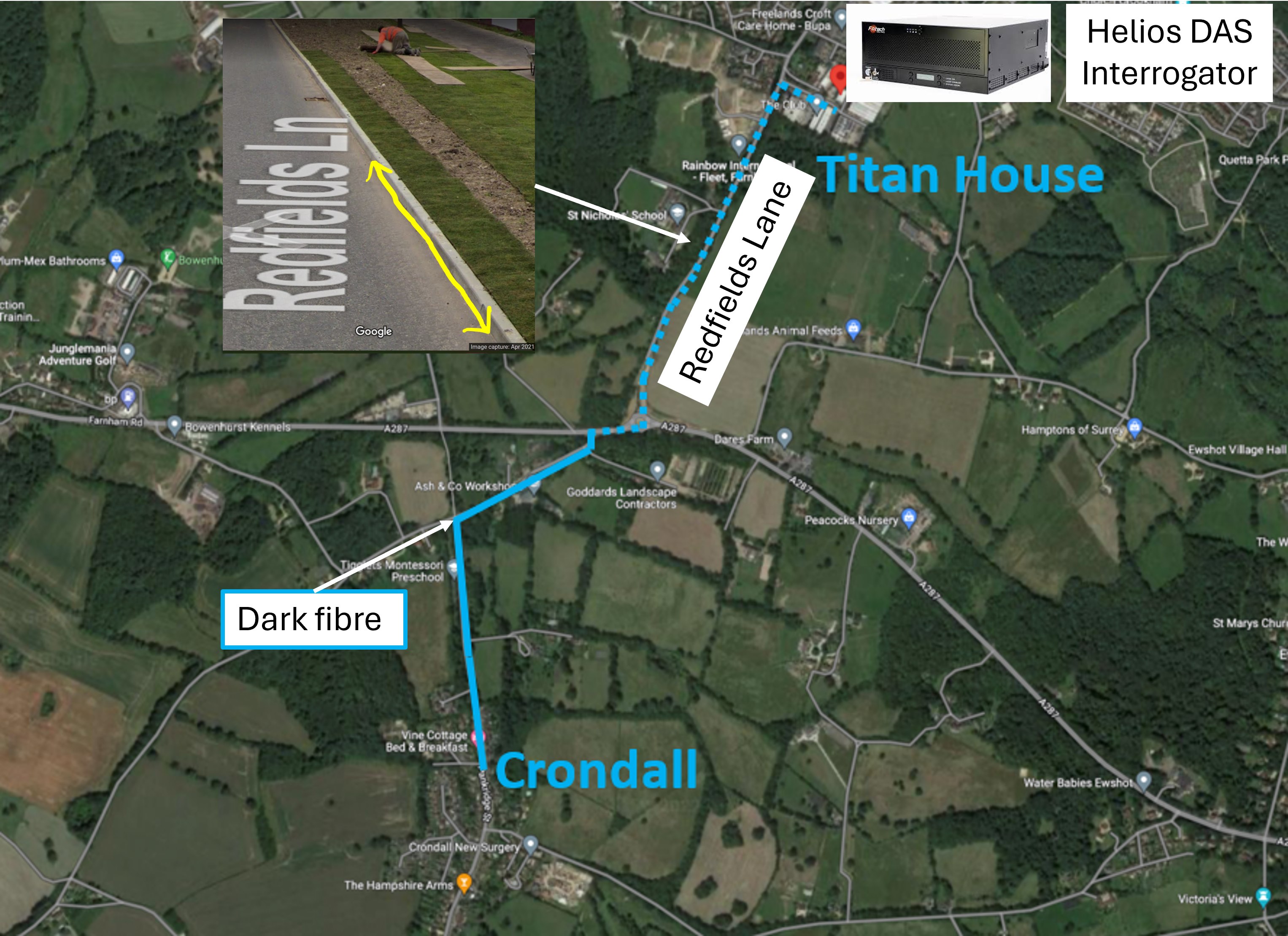}
        \label{fig:subfig_a}
    \end{subfigure}

    \vspace{0.3cm}

    \begin{subfigure}[b]{0.45\textwidth}
        \centering
        \includegraphics[width=0.9\linewidth]{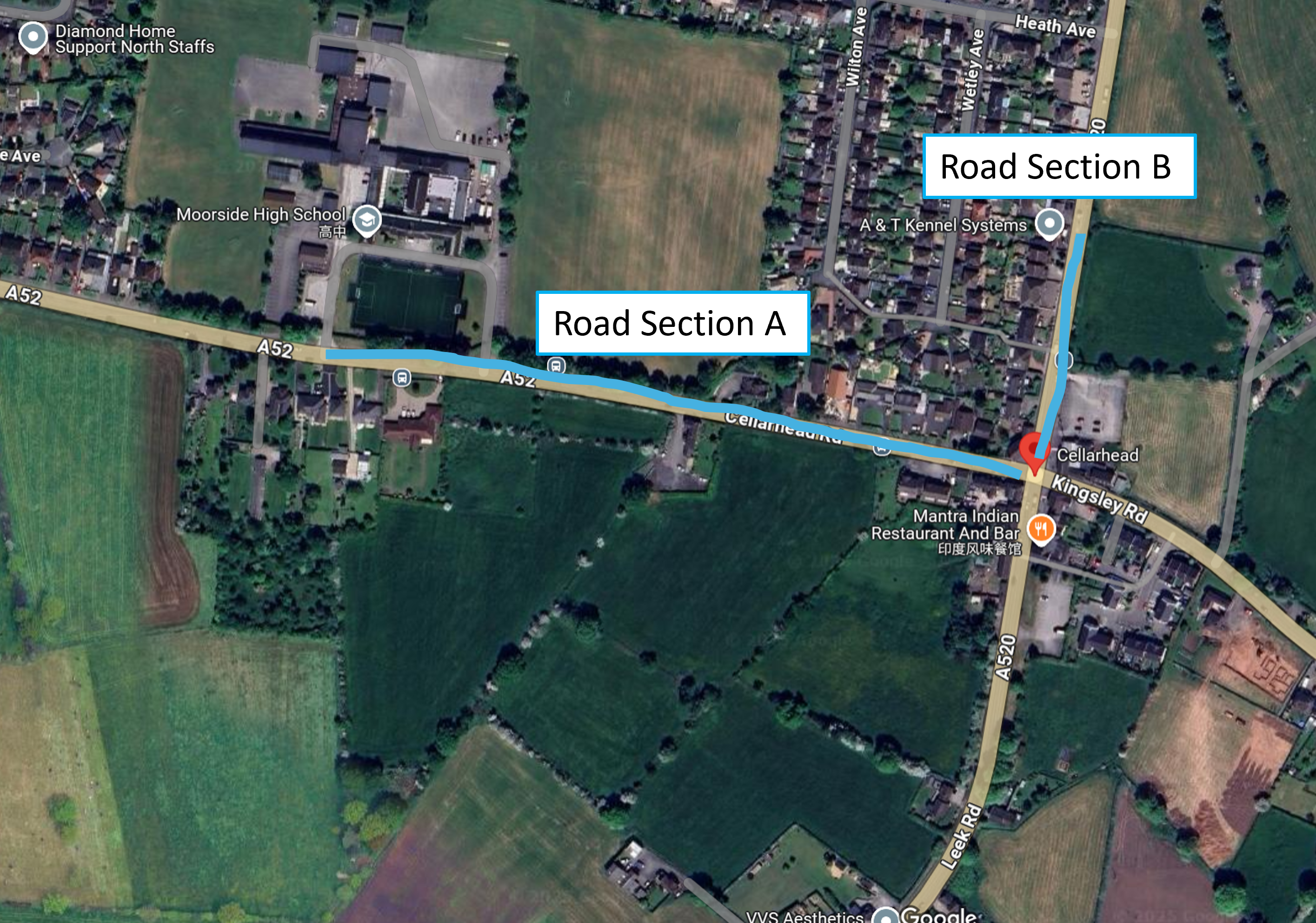}
        \label{fig:subfig_b}
    \end{subfigure}

    \caption{(a) Experiment field in Redfields Lane. (b) Experiment field in Cellarhead Junction.}
    \label{fig:meta_learning}
\end{figure}

\begin{figure*}[t!]
     \centering
     \begin{subfigure}[b]{0.48\textwidth}
         \centering
         \includegraphics[width=\textwidth]{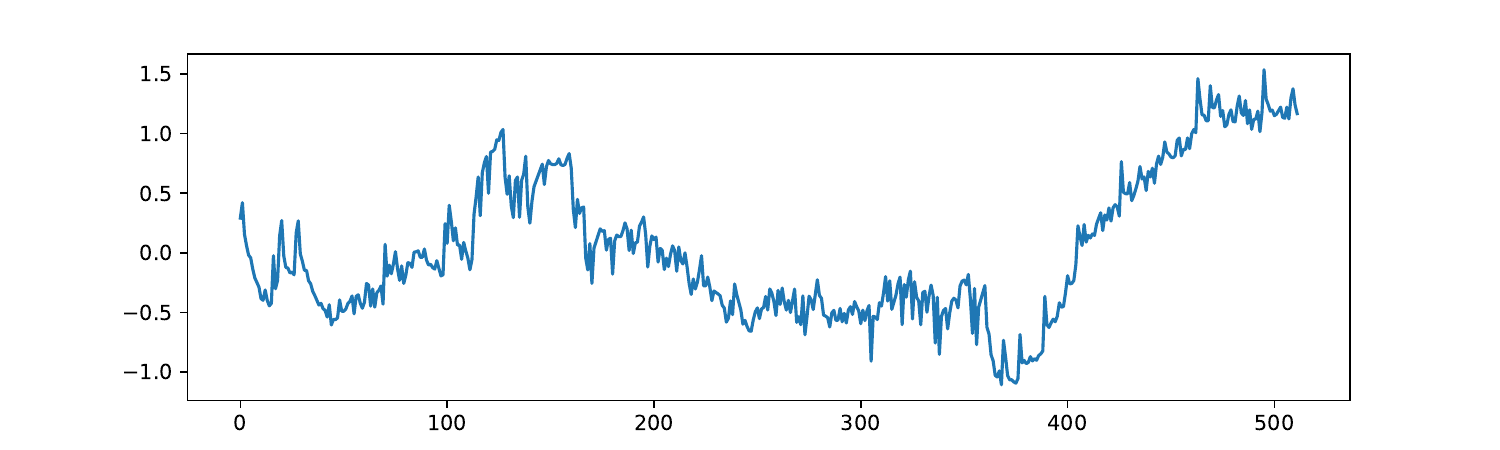}
         \caption{Time sequence walking signal (Redfields Lane)}
         \label{fig:y equals x}
     \end{subfigure}
     \hfill
     \begin{subfigure}[b]{0.48\textwidth}
         \centering
         \includegraphics[width=\textwidth]{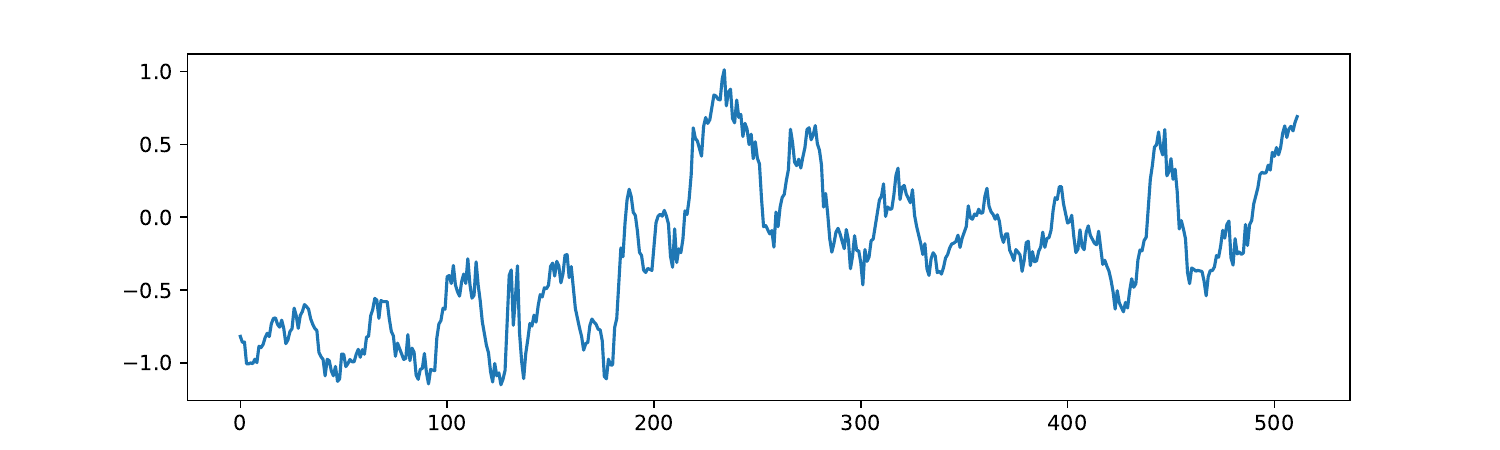}
         \caption{Time sequence walking signal (Cellarhead Junction)}
         \label{fig:three sin x}
     \end{subfigure}
     \hfill
     \begin{subfigure}[b]{0.48\textwidth}
         \centering
         \includegraphics[width=\textwidth]{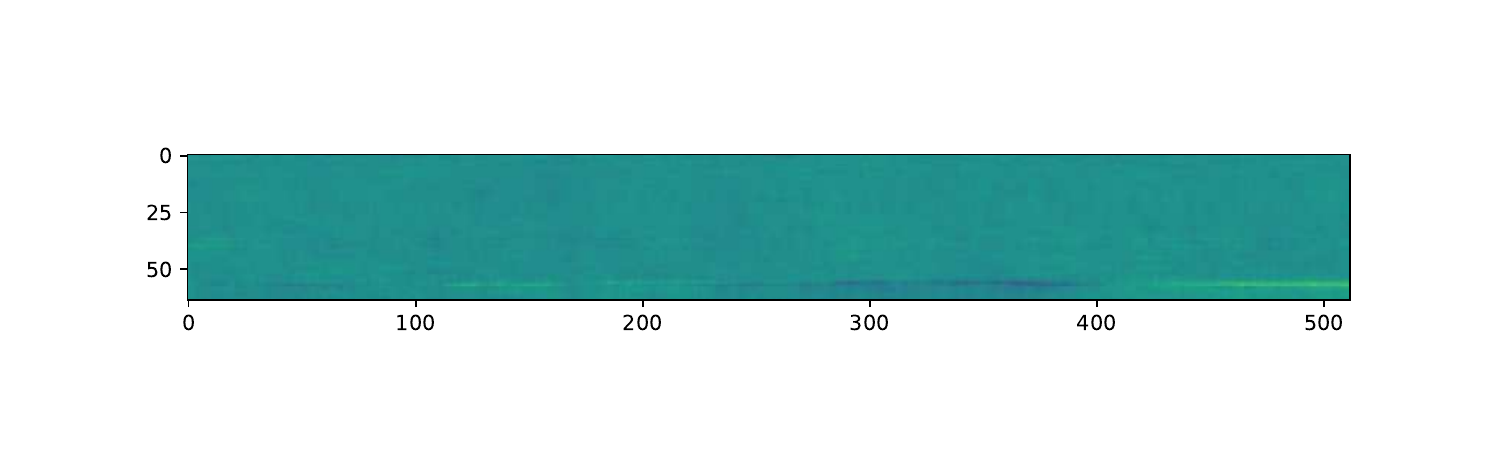}
         \caption{Window sampled walking signal (Redfields Lane)}
         \label{fig:five over x}
     \end{subfigure}
     \hfill
     \begin{subfigure}[b]{0.48\textwidth}
         \centering
         \includegraphics[width=\textwidth]{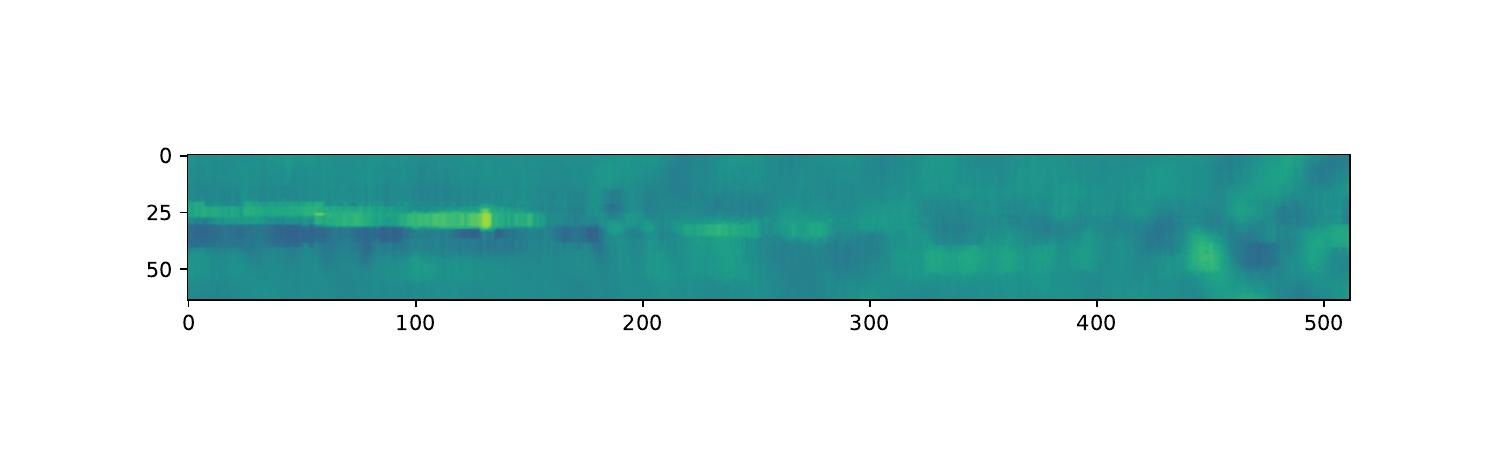}
         \caption{Window sampled walking signal (Cellarhead Junction)}
         \label{fig:five over x}
     \end{subfigure}
     \hfill
     \begin{subfigure}[b]{0.48\textwidth}
         \centering
         \includegraphics[width=\textwidth]{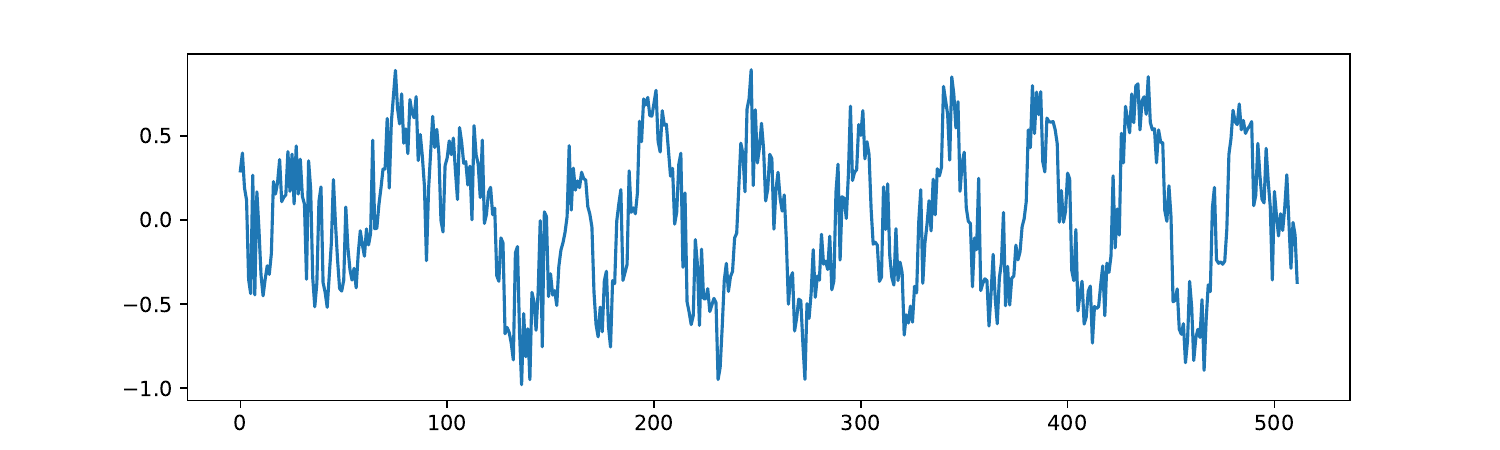}
         \caption{Time sequence cycling signal (Redfields Lane)}
         \label{fig:five over x}
     \end{subfigure}
     \hfill
     \begin{subfigure}[b]{0.48\textwidth}
         \centering
         \includegraphics[width=\textwidth]{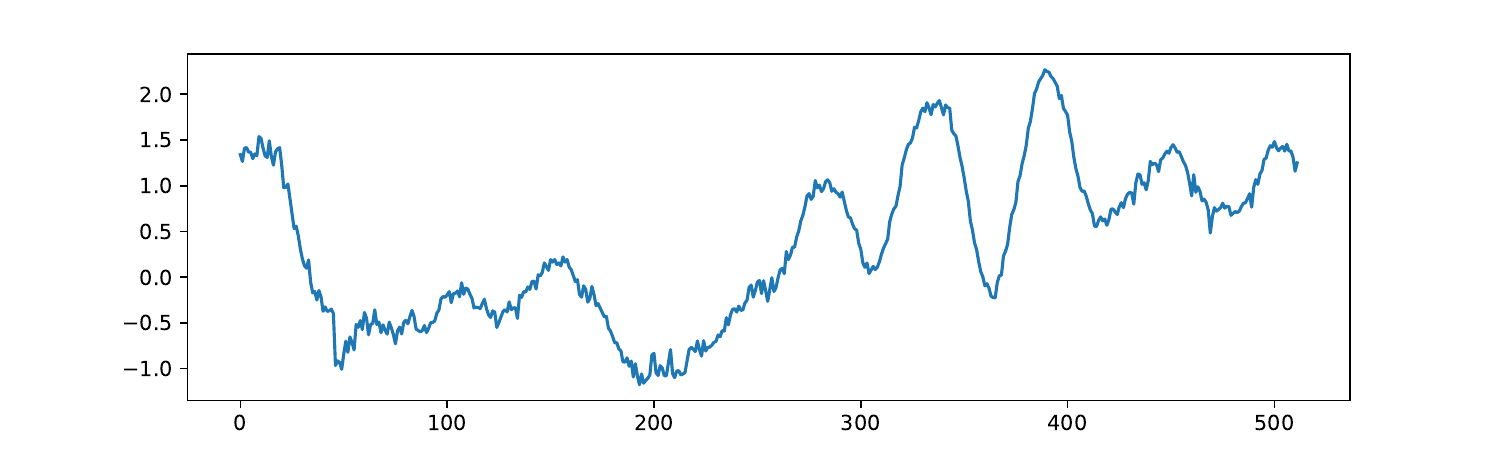}
         \caption{Time sequence cycling signal (Cellarhead Junction)}
         \label{fig:five over x}
     \end{subfigure}
     \hfill
     \begin{subfigure}[b]{0.48\textwidth}
         \centering
         \includegraphics[width=\textwidth]{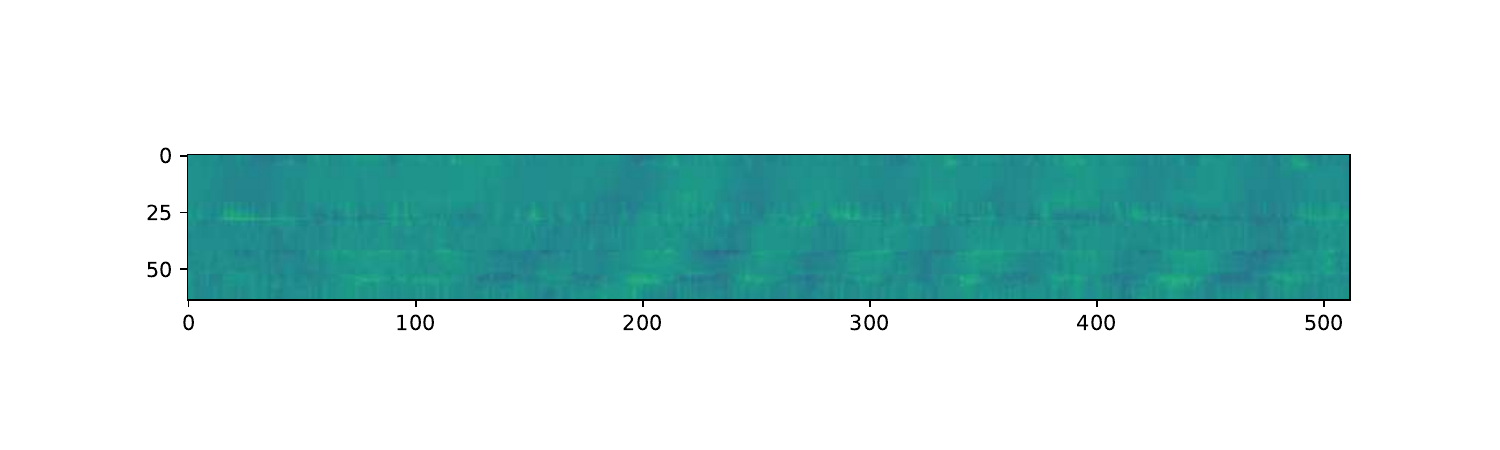}
         \caption{Window sampled cycling signal (Redfields Lane)}
         \label{fig:five over x}
     \end{subfigure}
     \hfill
     \begin{subfigure}[b]{0.48\textwidth}
         \centering
         \includegraphics[width=\textwidth]{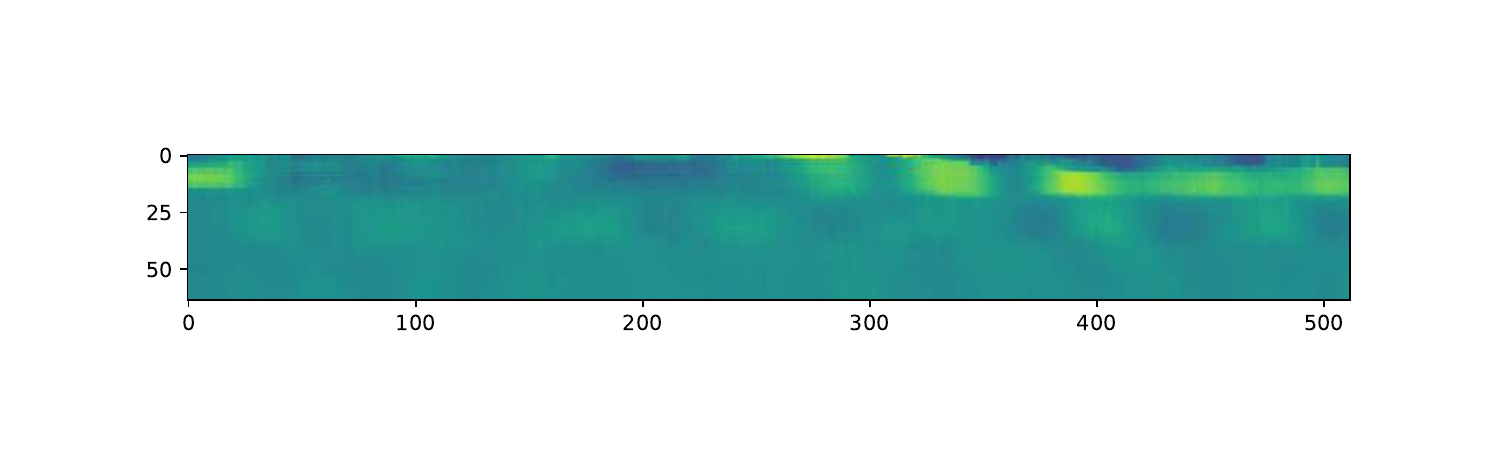}
         \caption{Window sampled cycling signal (Cellarhead Junction)}
         \label{fig:five over x}
     \end{subfigure}
\caption{The examples of processed DAS data from Redfields Lane and Cellarhead Junction.}
\label{fig:1D2D}
\end{figure*}

We conducted another experiment using the same DAS interrogator and data collection approach in Cellarhead Junction, which is located in Werrington, a city in the north of Birmingham, and it is geographically different from Redfields Lane. For the experiment at Cellarhead Junction, we collected data on two different sections of the junction. We collected data from the two highlighted sections of the east-west road A52 and north-south road A520 as shown in Fig. 3(b), and we named them Cellarhead Junction Road Section A and Cellarhead Junction Road Section B. We collected the two data sets from A52 and A520 for studying the \textbf{DR} scenario. There are several differences between the experiments in Redfields Lane and Cellarhead Junction, which can be summarized as follows.

\subsubsection{Climate differences}
Since Cellarhead Junction is in central England and Redfields Lane is in the south of England, there are relatively differences in climate between them. On the one hand, the annual precipitation of the Redfields Lane is 881 mm, and the annual average temperature is 10.4 °C. On the other hand, the annual precipitation and the average temperature of Cellarhead Junction are 805 mm and 9.4 °C, respectively.

\subsubsection{Weather conditions}
Both of the data collection experiments were completed during the day time, starting at around 10 am and ending at 4 pm the same day. The data for Redfields Lane was collected in July 2022, and the weather was clear and dry. The experiment at Cellarhead Junction was conducted in March 2022, and the weather was cloudy during the data collection.

\subsubsection{Fiber conditions}
The fiber used for the Redfields Lane data collection was dark fiber, but the fiber at Cellarhead Junction was purposely laid, which is a brand-new fiber with good condition.

\subsubsection{Deployment differences}
The Cellarhead Junction experiment's optical fibers were located 50cm underground and it was sheathed. In contrast, in the case of Redfields Lane, the optical fibers were located 20cm underground, with no additional sheathing. From the perspective of relative position to the road,  the fiber at Cellarhed Junction is farther away from the edge of the road.

\subsubsection{Sampling rate}
Two different DAS sampling rates were used in experiments. The DAS signal sampling rate in the experiment of Redfields Lane was 500 Hz, while the DAS sampling rate in Cellarhead Junction was 750 Hz.

After completing the above experiments, we have obtained three separate data sets, which are from Redfields Lane, Cellarhead Junction Road Section A, and Cellarhead Junction Road Section B. The difference in environmental conditions between the two road sections is smaller than that between the two regions. These data sets constitute the raw materials for our study of the three cases of \textbf{SD}, \textbf{DA}, and \textbf{DR}.

\subsection{Pre-process of the data}

In our experiments, we collected three types of data: DAS signals, GPS trajectories, and video recordings. The video was employed to assist in recording the activities of the subjects, while the GPS data provided the real-time location information of the subjects.   After synchronizing all data along the time axis, we utilized the given GPS position and timestamps to sample the DAS signal segment $\mathbf{X}$ among the continuous DAS signal, effectively associating specific DAS measurements with the movement of individual participants. Therefore, the preprocessing of raw DAS data can be summarized into the following four steps.

\begin{itemize}
\item \textbf{Timing synchronization:}  We synchronize the timeline of each segment of DAS data, GPS data, and video data to establish a corresponding relationship between the data. After the corresponding relationship is given, when the mode of data is required, it can be indexed by the timestamp.
\item \textbf{Window sampling:} According to the user location in GPS and the given longitude and latitude of DAS fiber deployment, we can query the relative position of the user and the fiber at any time, and calculate the nearest bin to the subject. After determining both the time step and the location of the bin, we set a sampling window of $H=64, W=512$ to obtain the two-dimensional tensor data $\mathbf{X}_{b,t}$ as described in Eq \eqref{eq2}.
\item \textbf{Data cleaning:} Since GPS may have positioning errors in practice, the sampling window sometimes fail to capture the signal recording the active traveler's motion. In order to prevent invalid data from affecting the training results, we set a phase extreme value threshold $x_\text{th}$. As described in \eqref{eqth}, if the absolute value of the extreme value of the signal in the sampling window is not greater than the threshold, the data sample is considered invalid.
\begin{equation}
\text{Validity}(\mathbf{X}) =
\begin{cases}
\text{Valid}, & \text{if } \max\left( |\mathbf{X}| \right) > x_{\text{th}}, \\
\text{Invalid}, & \text{if } \max\left( |\mathbf{X}| \right) \leq x_{\text{th}}.
\end{cases}\label{eqth}
\end{equation}
\item \textbf{Data labeling:} For each DAS data segment obtained by window sampling, we add labels according to the active travel behavior recorded in the video, where $y = 0$ represents walking and $y = 1$ represents cycling.
\end{itemize}

After preprocessing the data collected from two regions, we obtained three datasets, and we named them \textit{Redfields Lane (Red) dataset}, \textit{Cellarhead Junction A (CA) dataset}, and \textit{Cellarhead Junction B (CB) dataset}. Due to the different experimental durations, the number of data samples obtained varies greatly, where they contain 1085, 122, and 126 valid data samples, respectively. The samples of each dataset are randomly divided into training data and test data. For the \textit{Red dataset}, the division ratio is 9.4:0.6, while the division ratio of the other two datasets is 8:2 to ensure sufficient test data.

The examples of valid DAS data samples of the same subject are plotted in Fig. \ref{fig:1D2D}, where Fig. \ref{fig:1D2D}(c)(d)(g)(h) are the data samples for walking and cycling from both experiments.  Fig. \ref{fig:1D2D}(a)(b)(e)(f) plot a time domain signal in the sampling window as auxiliary examples. Observing any pair of samples of an activity, it can be found that even the same motion has significant differences in terms of signal shape, frequency, and amplitude in different environments. Taking the maximum amplitude of cycling as an example, the maximum amplitude of the data in the \textit{Red dataset} is less than 1, while this number can reach 2 in the data collected from Cellarhead Junction. These significant differences can explain why the classification accuracy drops dramatically when the training data and test data come from different sources, as we observed in Fig. \ref{Fig.2}, further demonstrating that it is necessary to invoke novel technical approaches to resolve the generalization problem.

\section{Solution I: FL-enabled IoT Solution}

In order to solve the problem of the DL model trained on the local DAS dataset not having sufficient scalability and generalization ability to migrate to other DAS nodes, we proposed a solution based on FL and IoT network. The core principle of this solution is to use the IoT network to link multiple DAS nodes to produce a generalized model using FL, and it enables geographically dispersed DAS nodes to collaborate in training shared global models.

\begin{figure*}[t]
\centering
\includegraphics[width=0.8\textwidth]{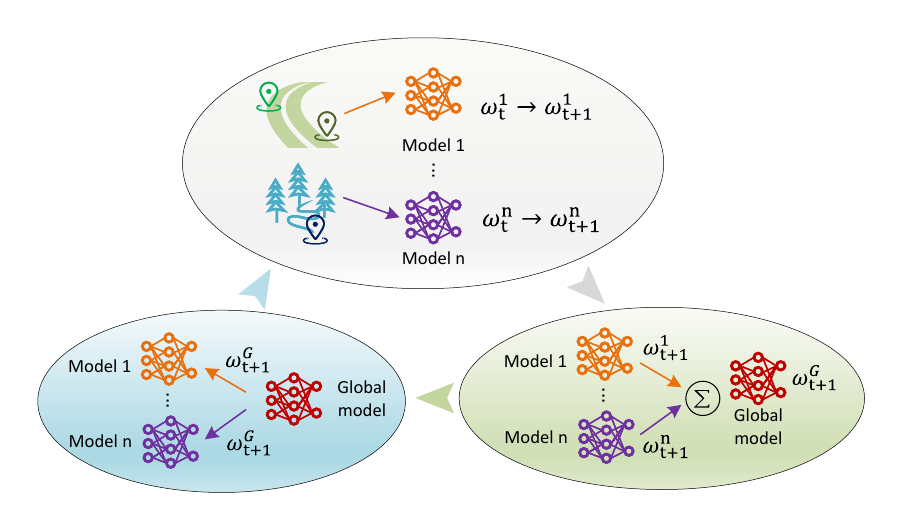}
\caption{Work flow of the FL enabled IoT Solution for DAS system.}
\label{Fig.FL}
\end{figure*}

\subsection{Why FL}

The traditional centralized learning paradigm with IoT assistance is certainly a possible solution. This solution requires the data collected from all DAS interrogators to be transmitted and aggregated at a central server for model training. However, this solution encounters the following challenges in DAS practice. First, network bandwidth limitations make it impractical to transmit a high volume of high-resolution DAS signals. In practice, DAS coverage can be tens of kilometres, and each DAS node collects an enormous number of DAS samples. It is not an economical solution to transmit all the data. Moreover, this solution may still face the dilemma of unbalanced data samples. The number of active travelers may vary significantly on different sections of the road. As a result, some sections receive a large amount of data, while some sections have very limited data samples. After mixing these data and performing uniform sampling during training, the large-scale sample data will dominate the training, resulting in the deterioration of the model performance on DAS nodes with low popularity.

FL allows each DAS node to perform local model updates and only share model parameters (such as gradients or weights) with the central aggregator. This approach inherently improves privacy and reduces communication overhead, especially for the lightweight model SR-Net (18Mb). The transmission of this model saves significant communication resources compared to the transmission of a large amount of training data. In addition, the model weight coefficients of the FL federated aggregation can be flexibly adjusted to avoid the problem of unbalanced data.

\subsection{Work Flow}

As shown in Fig. \ref{Fig.FL}, in the context of the DAS system, FL offers a practical framework to collaboratively train the SR-Net model across multiple geographically distributed DAS nodes sharing the parameters of the model. Each \textbf{DAS node} (e.g., Redfields Lane, Cellarhead Junction Road Section A, and Cellarhead Junction Road Section B) acts as an independent sensing and learning entity. These nodes collect raw vibration signals caused by activities through underground fiber optics, pre-process the signals locally into the standard shape of two-dimensional feature matrices $\mathbf{X}^m_{b,t}$ as per Eq.~(2), where $m$ represents the index of the DAS node. Then, the FL process for DAS operates is given as follows

\begin{enumerate}
    \item \textbf{Model Initialization:} At each DAS node, a SR-Net model $\omega^m_0$ is initialized, and a global SR-Net model $\omega^G_0$ is initialized at the cloud server. This model can be either pre-trained or initialized randomly.

    \item \textbf{Local Model Update:} Each DAS node $m$ trains the SR-Net model locally on its own dataset $D^{m} = \{(\mathbf{X}^{n,m}, y^{n,m})\}$ using the local optimizer to obtain trained model $\omega^m_t$. The local optimization problem at each node is formulated as:
    \begin{equation}
        \min_{\omega_m} \frac{1}{N_m} \sum_{n=1}^{N_m} \mathcal{L}(f_{\omega_m}(\mathbf{X}^{n,m}), y^{n,m}),
    \end{equation}
    and the loss function $ \mathcal{L}()$ is given by
    \begin{equation}
        \mathcal{L}(\hat{y}, y) = -\left[ y \cdot \log(\hat{y}) + (1 - y) \cdot \log(1 - \hat{y}) \right],
    \end{equation}\label{loss}
    where $\hat{y} = f_{\omega}(\mathbf{X})$ is the model's predicted probability of class.

    \item \textbf{Model Upload:} After several local epochs of local model updates, each node transmits the updated model parameters $\omega^m_t$ to the central server. AS the DAS signals are not transmitted, the data privacy of the DAS node (e.g. location) is further protected.

    \item \textbf{Model Aggregation:} The server aggregates the received parameters using the federated average approach given by
    \begin{equation}
        \omega^{\text{G}}_t = \sum_{m=1}^{M} \frac{1}{M} \omega^m_t.
    \end{equation}

    It is worth noting that the federated aggregation scheme is flexible. In practice, advanced aggregation schemes based on data volume or specific weight strategies can also be employed.

    \item \textbf{Global Model Distribution:} The updated global model $\omega^G_{t}$ is redistributed to each DAS node for the next round of training, progressively refining the model's generalization across diverse environments.
\end{enumerate}

By employing the given FL framework, we enable active travel classification models to collaboratively learn from distributed DAS nodes while preserving data privacy, enhancing transmission efficiency, and achieving superior cross-location generalization compared to independently trained models.

\begin{figure*}[t]
\centering
\includegraphics[width=0.7\textwidth]{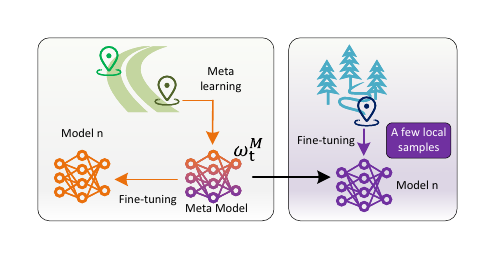}
\caption{Work flow of meta learning based initialization.}
\label{Fig.10}
\end{figure*}

\subsection{Critical Discussion}
Notwithstanding the proposed FL-based IoT solution's strengths, it is imperative to address the inherent limitations of FL in the DAS system. On the one hand, despite the fact that FL serves to reduce the demand for the transmission of raw DAS data, the model updates nevertheless continue to rely heavily on reliable data links to DAS nodes. In practical deployments, it has to be noted that not all DAS nodes' networks are online during every round of federation aggregation. Uplink and downlink link congestion have the potential to impede the contribution of knowledge to the global model by certain nodes, as well as the downloading of the latest version of the global model. It is anticipated that long-term disparities in network availability will result in a global model that is biased towards active nodes with richer datasets. On the other hand, using the global model aggregation, certain nodes having strong local environmental bias are likely to exhibit suboptimal performance as the global model is averaged to fit the general DAS cases. Without adaptation or fine-tuning at the specific node, the FL framework risks averaging out unique signal characteristics at a specific DAS node that are critical for accurate classification.

\section{Solution II: Meta learning based initialization}

While FL is an effective solution for collaborative model training, it still requires well-connected network conditions and multiple rounds of parameter exchange. In order to address the challenge of model generalization for DAS nodes without network resource assistance, we proposed a meta-learning-based initialization scheme. This scheme aims to train an initialized model with high generalization potential using training samples collected in other regions. Following the deployment of the generalized model to a particular DAS node, the subsequent process entails only a quick fine-tuning process of the model to facilitate adaptation to the environment and task requirements of the designated target area.

\subsection{Why Meta-learning}
Although there are a number of technologies dedicated to model initialization and retraining (e.g., transfer learning), these technologies often require a large amount of local data and computing power. Unfortunately, these retraining or fine-tuning approaches are often impractical in DAS scenarios. First, it may take a long time for the DAS node to collect and label enough local DAS data to support retraining. Second, retraining may require a large number of training rounds, which is a challenge for the DAS node's computing power. Consequently, there is a pressing demand for learning techniques that enable low-complexity adaptation to new environments using only a few local samples.

To address this, we invoke a Reptile algorithm, which aims to train a general model that can quickly adapt to new tasks with minimal data. Among available methods, Reptile offers an ideal balance of simplicity and effectiveness for DAS systems. First, Reptile is a kind of model-agnostic meta-learning (MAML) method that can be directly used to extend SR-Net to a meta learner, which perfectly fits our scenario. In addition, it only requires a few local gradient steps and a few local data samples for local fine-tuning. Finally, as a first-order gradient-based method, Reptile avoids the high computational cost of second-order optimization used in the original MAML, making it more suitable for DAS nodes. These properties make Reptile particularly appealing for DAS environments, enabling robust activity recognition even in isolated deployments with limited local data.

\subsection{Reptile-Based Model Adaptation for DAS Nodes}

Reptile is a gradient-based meta-learning method that learns a model initialization capable of adapting quickly to new tasks using only a few training samples and gradient steps. In the considered scenario, each task corresponds to an activity classification problem defined on a specific DAS node, characterized by its unique factors caused by the aforementioned local DAS deployment and environment. The Reptile algorithm for the DAS system is divided into three main phases, namely meta-model training, local fine-tuning, and testing.

\subsubsection{Meta-Training Phase}

The goal of the meta training phase is to obtain a shared model initialization $\omega^M_t$ that enables fast adaptation to any new DAS nodes. The meta training phase physically takes place at the DAS node with rich samples or on a cloud server having DAS data collected from multiple DAS nodes. Following the standard Reptile algorithm, it performs multiple rounds of tasks sampling and training based on previously collected datasets $\mathcal{D}_{n \neq m}$ from other DAS nodes. In our experiment, the \textit{Red data set} played the role of the $\mathcal{D}_{n}$ as it has significantly richer data samples.

Each meta-task $\tau$ is also defined as a binary classification problem. For each task, we randomly sample a support set $\mathcal{D}^{\text{train}}_{n}$ and a query set $\mathcal{D}_n^{\text{test}}$. The SR-Net model is then trained on $\mathcal{D}^{\text{train}}_{n}$ for $k$ steps using the Adam optimizer following the binary cross-entropy loss described in Eq. \eqref{loss}.

Then, starting from the previous model $\omega_{t-1}^M$ (or initialized model $\omega_0^M$) , we sample several batch of training data from $\mathcal{D}^{\text{train}}_n$, and perform $k$ steps of gradient descent on  $\omega_{t-1}^M$, producing a task-specific parameter $\omega'$:
\begin{equation}
\omega^n_t = \text{Adam}_k(\omega_{t-1}^M, \mathcal{D}_\tau^{\text{train}}), \label{NT}
\end{equation}
where the model $\omega^n_t$ is adapted to task $\tau$. Then, we update the meta model following
\begin{equation}
\omega^M_t \leftarrow \omega_{t-1}^M + \epsilon (\omega^n_t - \omega_{t-1}^M), \label{meta}
\end{equation}
where $\epsilon$ is the learning rate of meta step and the model $\omega^M_t$ is the desired meta model. This procedure is repeated over serious tasks sampled from the support data set, allowing $\omega^M_t$ to become a versatile starting point for new DAS deployments.

\subsubsection{Local Fine-Tuning}

After the meta-training phase is complete, the meta model initialization $\omega^M_t$ is deployed to a new DAS node as its initialized model (e.g., Cellarhead Junction Road Section A and Cellarhead Junction Road Section B), which only requires using a small number of local samples $\mathcal{D}^{\text{train}}_{m}$ (e.g., 5 to 10) to perform a few-shot learning as fine-tuning. The gradient updates of the model using local support sets is given by
\begin{equation}
\omega^m_t = \omega^M_t - \alpha \nabla_\omega \mathcal{L}(f_\omega(\mathbf{X}_m), y), \label{FT}
\end{equation}
where $\alpha$ is the learning rate of fine-tuning. This fine-tuning process is computationally lightweight and can be performed entirely offline without requiring any communication with a central server. As a result, it is well-fitted for resource-constrained or intermittently connected DAS nodes.

\subsubsection{Testing}

In the final phase, the adapted model $\omega^m_t$ is evaluated on the test data set $\mathcal{D}^{\text{test}}_{m}$ at the new DAS node. This step emulates real-world deployment, where the model must operate reliably with minimal calibration data. The full workflow of the Reptile assisted DAS is given in \textbf{Algorithm 1}.

\begin{algorithm}
\caption{Reptile-Based Few-Shot Adaptation for DAS}
\begin{algorithmic}[1]
\STATE \textbf{1) Meta-Training Phase:}
\STATE Initialize $\omega^M_0$, the vector of initial model parameters
\FOR{iteration $= 1, 2, \dots$}
    \STATE Sample task $\tau$ from $\mathcal{D}_{n \neq m}$
    \STATE Train task-specific model following Eq. \eqref{loss} and Eq. \eqref{NT}
    \STATE Update meta model $\omega^M_t$ following Eq. \eqref{meta}
\ENDFOR

\STATE \textbf{2) Local Fine-Tuning Phase:}
\STATE Initialize local model $\omega^m_0$ following $\omega^M_t$
\STATE Sample local support set $\mathcal{D}^{\text{train}}_{m}$
\FOR{training shot $= 1, 2, \dots$}
\STATE Update local model following Eq. \eqref{FT}  to get $\omega^m_t$
\ENDFOR

\STATE \textbf{3) Testing Phase:}
\STATE Evaluate adapted model $\omega^m_t$ on test set $\mathcal{D}^{\text{test}}_m$.
\end{algorithmic}
\end{algorithm}

\subsection{Critical Discussion}
While the proposed Reptile approach demonstrates promising potential for rapidly adapting activity classification models for DAS nodes, there are still some limitations to this approach.

The current framework assumes that the domain shift between DAS nodes is in space. However, in practical scenarios, temporal variations are another factor changing the environment of the DAS node. Time-varying factors such as ambient temperature, precipitation, and seasonal changes can subtly or significantly change the vibration propagation characteristics of DAS signals. Since the proposed meta-learning approach only offers an initialized model, it lacks the capability of updating against long-term or time-varying conditions. This limits its robustness in dynamic environments where signal characteristics evolve slowly or cyclically.

The other limitation is that the evaluation of the proposed approach is limited by the conditions of the experiment and the collected data. While we focus on binary activity classification (walking vs. cycling), the effectiveness of Reptile in handling more complex tasks such as multi-class activity recognition, object detection in high-traffic areas, or signal attenuation due to extreme weather remains unexplored. These scenarios may challenge the adaptability and stability of meta-learning based initialization. Our future experiments will explore these more challenging tasks by expanding data collection and diversifying task objectives.

\section{Experiment Results}

\begin{figure}[t]
\centering
\includegraphics[width=0.5\textwidth]{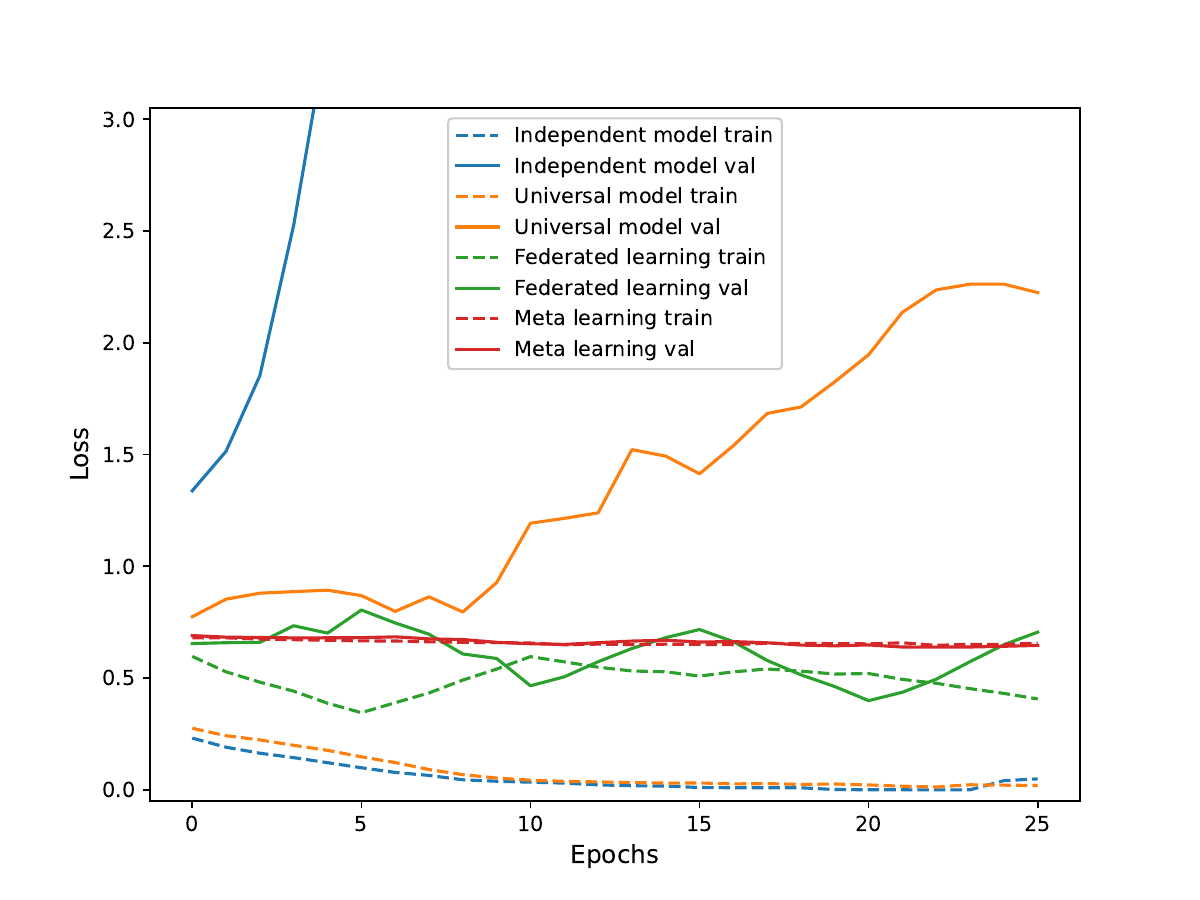}
\caption{Training and validation loss in \textbf{DA} case.}
\label{Fig.7}
\end{figure}

In this section, we present the performance of the proposed two generalized DAS using three DAS datasets of \textit{Red,}  \textit{CA}, and \textit{CB}. We consider both cross-node and within-node evaluations to assess the adaptability of in realistic deployment scenarios.
We considered two baseline strategies of \textbf{Independent model} and \textbf{Universal model} to validate the generalization performance. The specific descriptions of the considered cases are described as follows.

\begin{itemize}
    \item \textbf{Independent model:} This scenario simulates an independent DAS node, and the SR-Net model is trained with local data and tested with another dataset. In this case, the training and test datasets are sampled from \textit{Red} and the test dataset is sampled from \textit{CA}.
    \item \textbf{Universal model:} A universal model is assumed to be trained and tested with mixed data collected from all nodes, including \textit {Red, CA} and \textit {CB}.
    \item \textbf{Federated learning:} There are two or three DAS nodes in the simulation cases of \textbf{DA} and \textbf{DR}. In the case of \textbf{DA}, we assume there are 3 nodes at \textit {Red, CA} and \textit {CB}. They use their data for local training. The latest aggregate model is tested separately on the testing datasets $\mathcal{D}_\text{Red}^\text{test}$, $\mathcal{D}_\text{CA}^\text{test}$, and $\mathcal{D}_\text{CB}^\text{test}$. In the case of \textbf{DR}, we assume there are 2 nodes at \textit{CA, and CB}, and the simulation process follows the same logic as the \textbf{DA} case.
    \item \textbf{Meta-learning:} A generalized initialization is trained using the dataset $\mathcal{D}_\text{Red}^\text{train}$, followed by few-shot local using data set $\mathcal{D}_\text{CA}^\text{train}$ and $\mathcal{D}_\text{CB}^\text{train}$. Finally, the trained local models are tested by $\mathcal{D}_\text{CA}^\text{test}$ and $\mathcal{D}_\text{CB}^\text{test}$.
\end{itemize}

Fig.~\ref{Fig.7} plots the training and validation loss of different approaches under the \textbf{DA} scenario, where the model is trained on the \textit{Red} dataset and validated on the \textit{CA} dataset.  This highlights the effectiveness of proposed generalization strategies in heterogeneous DAS environments. It can be found that in this case, although the training loss decreases, the validation loss of the independently trained model increases sharply. In contrast, the proposed scheme can obtain consistent training and validation losses, which indicates that the model is generalized for data samples from different datasets. It is worth noting that the meta-learning scheme always has a lower validation loss, while the validation loss in the FL scheme is reduced by federated aggregation but increased by local training.

\begin{figure}[htb]
\centering
\includegraphics[width=0.5\textwidth]{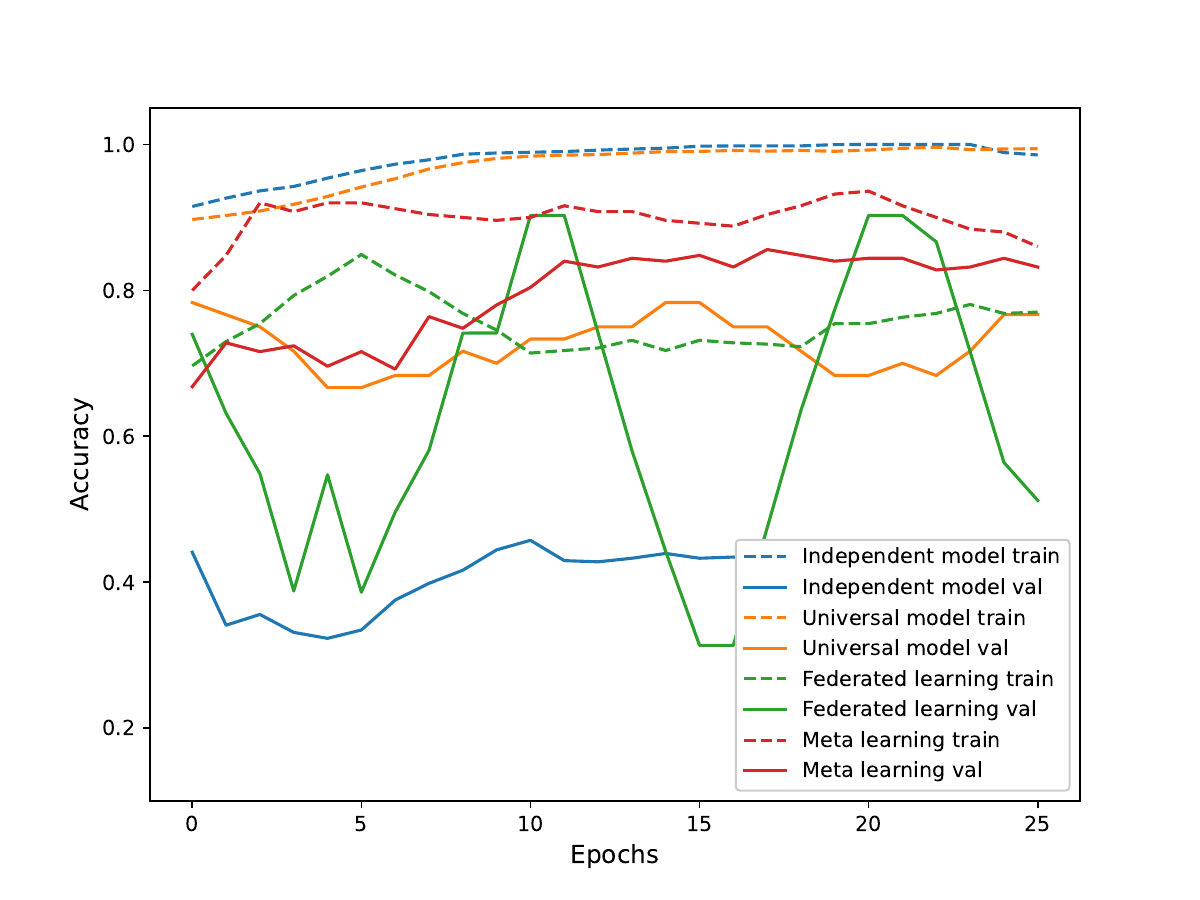}
\caption{Training and validation accuracy in \textbf{DA} case.}
\label{Fig.8}
\end{figure}

Fig.~\ref{Fig.8} plots the training and validation accuracy of different approaches under the \textbf{DA} scenario. The extremely low accuracy of the independent training clearly demonstrates that directly applying independently trained models to other DAS nodes is not feasible. Using all the data for training does achieve a promising result around 76\%, but it is slightly lower than meta-learning. The performance of FL shows extremely oscillating performance over the aggregation and local training of the model. The aggregation of the model occurs at 10 and 20 rounds. It can be observed that the performance of the model is significantly improved after each aggregation. This phenomenon is logical because using local data training destroys generality, while generality is re-implemented after each model aggregation.

\begin{figure}[htb]
\centering
\includegraphics[width=0.5\textwidth]{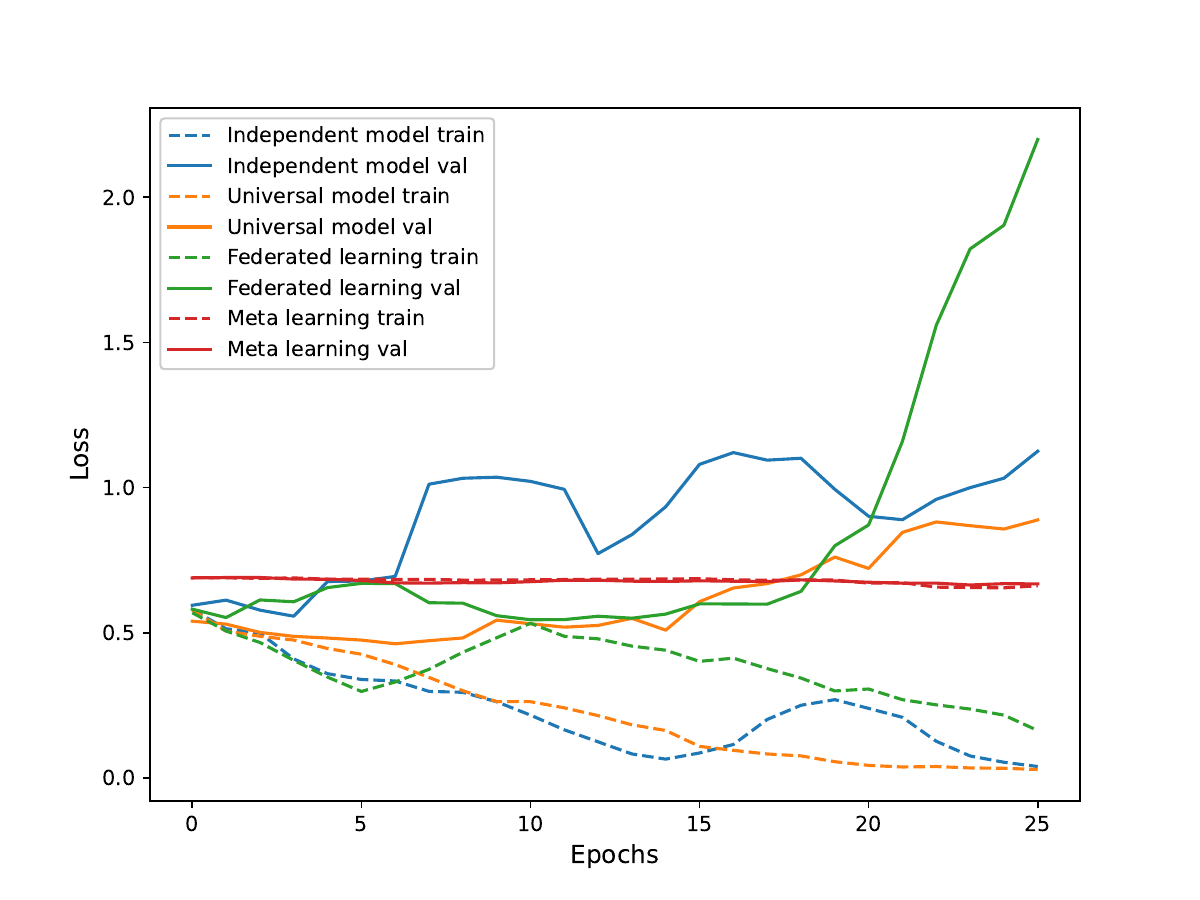}
\caption{Training and validation loss in \textbf{DR} case.}
\label{Fig.9}
\end{figure}

\begin{figure}[htb]
\centering
\includegraphics[width=0.5\textwidth]{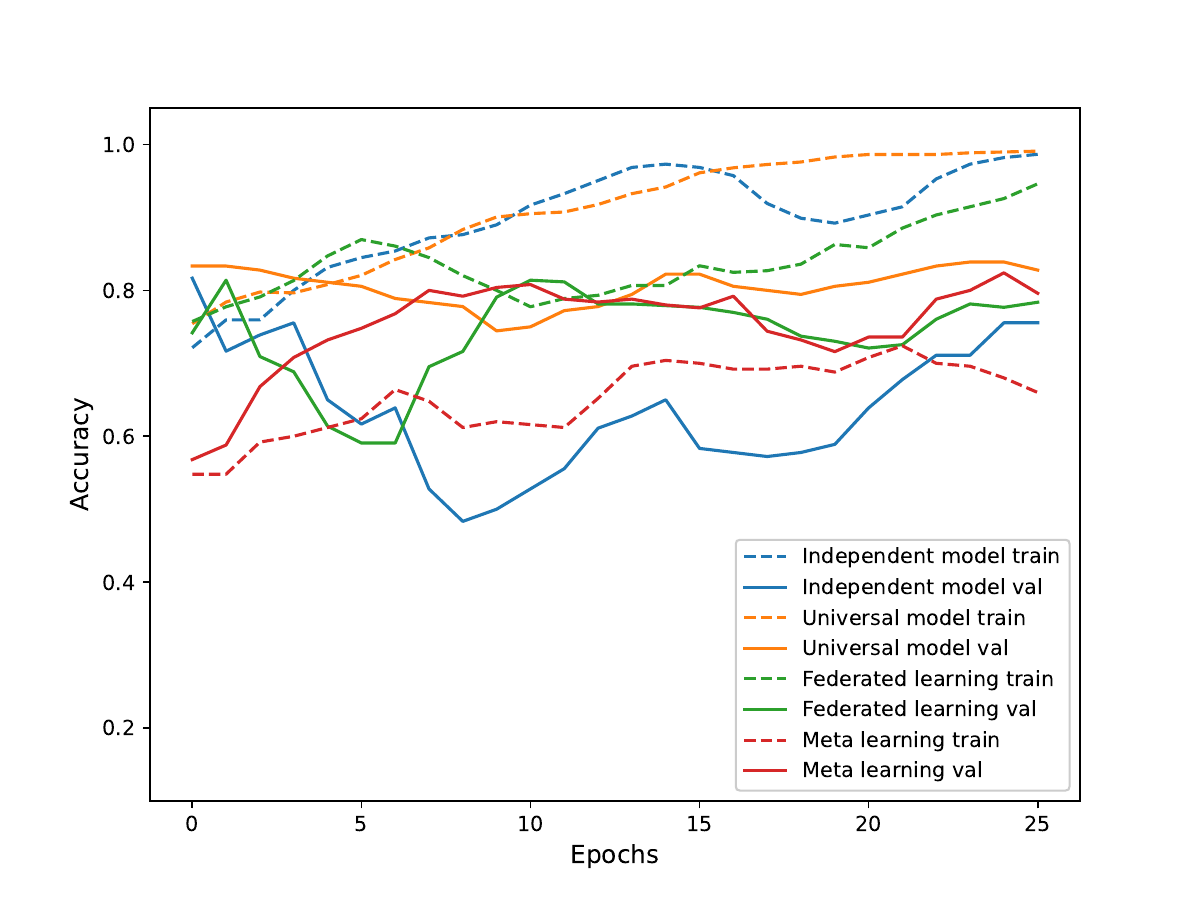}
\caption{Training and validation accuracy in \textbf{DR} case.}
\label{Fig.10}
\end{figure}

Fig.~\ref{Fig.9} shows the training loss curves for the \textbf{DR} case, where the model is trained on \textit{CB data set} and validated on \textit{CA data set}. All models converge steadily, and even independent training does not lead to a catastrophic increase in validation loss. This phenomenon suggests that the data similarity of DR is indeed significantly higher than the case of DA, which is in line with our expectations.

Fig.~\ref{Fig.10} illustrates the validation accuracy over epochs in the \textbf{DR} scenario. In this case, the performance of the solutions is not much different, concentrated in the range of 70-80\%.  In addition, due to the higher similarity in \textbf{DR} situations, the universal model has the best performance, and the performance gap between the model in the independent training mode and the model in the proposed solution is narrowed, but there is still a gap of 6\%. Therefore, the motivation to use FL or meta-learning in the \textbf{DR} case is weaker than in the \textbf{DA} case. However, it is worth noting that the FL solution can distribute the training across multiple hardware nodes, which is advantageous when the computing power of a single DAS node is limited.

\begin{figure}[htb]
\centering
\includegraphics[width=0.45\textwidth]{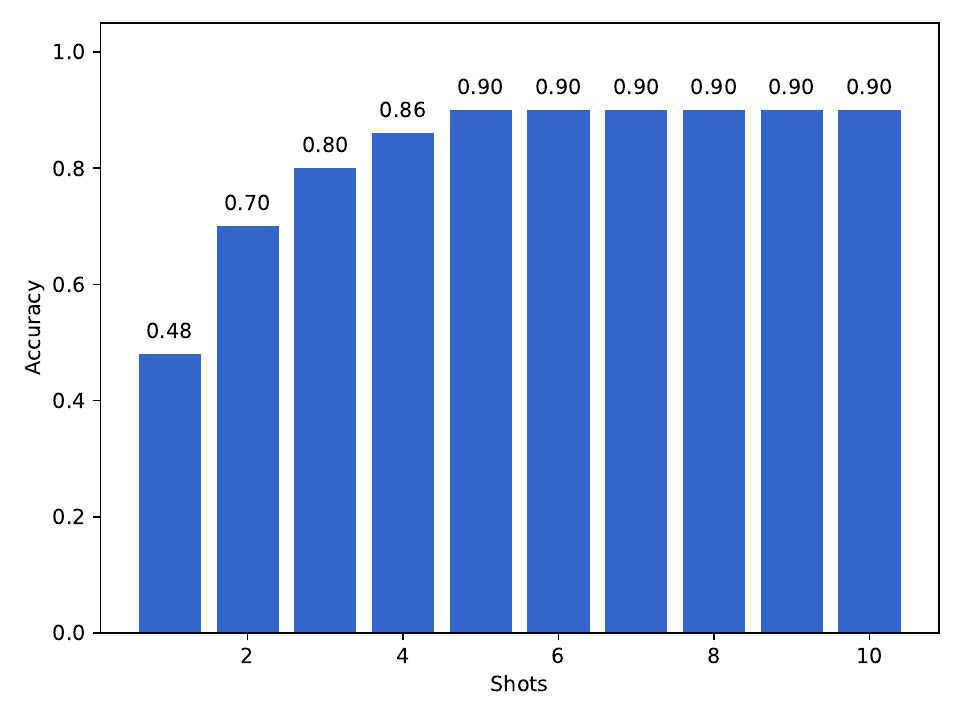}
\caption{Performance of the few-shot fine-tuning.}
\label{Fig.11}
\end{figure}

\begin{table}[]
\centering
\begin{tabular}{|c|c|c|c|c|}
\hline
Approach      & Case & Training & Test  & Test acc \\ \hline
Independent   & \textbf{DA}       & \textit{Red }           & \textit{CA }        & 0.3657   \\ \hline
Independent   & \textbf{DR}       & \textit{CB }            & \textit{CA }        & 0.7604   \\ \hline
Univeral      & \textbf{DA}       & \textit{All }     & \textit{Red  }      & 0.7425   \\ \hline
Univeral      & \textbf{DA}       & \textit{All  }    & \textit{CA  }       & 0.7917   \\ \hline
Univeral      & \textbf{DR}       & \textit{CA+CB}          & \textit{CA}         & 0.8368   \\ \hline
FL - 3Agents  & \textbf{DA}       & \textit{All }     & \textit{Red}        & 0.8452   \\ \hline
FL - 3Agents  & \textbf{DA}       & \textit{All}      & \textit{CA }        & 0.9051   \\ \hline
FL - 3Agents  & \textbf{DA}       & \textit{All }     & \textit{CB  }       & 0.8333   \\ \hline
FL - 2Agents  & \textbf{DR}       & \textit{CA+CB}          & \textit{CA}         & 0.9132   \\ \hline
FL - 2Agents  & \textbf{DR}       & \textit{CA+CB}          & \textit{CB }        & 0.8294   \\ \hline
Meta-learning & \textbf{DA}       & \textit{Red}            & \textit{CA }        & 0.86     \\ \hline
Meta-learning & \textbf{DR}       & \textit{CB}            & \textit{CA  }       & 0.9      \\ \hline
\end{tabular}
\caption{Test results of different generalization approaches.}
\label{tab:1}
\end{table}

In order to identify the gain obtained by the few-shot fine-tuning of the meta learning model. Figure~\ref{Fig.11} shows the performance during the few-shot adaptation of the meta-learned model on \textit{CA data set}. During the 10 shots,  with the number of support samples increased from 1 to 10 . The model's accuracy improves rapidly when feeding the beginning samples, reaching 90\% with 5 samples. The trend directly proves the gain obtained by few-shot fine-tuning, and it revealed that the optimal number of shots for the given DAS problem is 5. The increase also demonstrates that the pre-trained initialization enables efficient transfer and reliable performance in extremely low-sample cases.

Table 1 demonstrates the test results of different cases. The FL model and  Reptile model achieve a high accuracy, reaching over 90\% in both DA and DR cases. These results suggest that even in intra-site variations, collaborative and meta-learning frameworks provide strong generalization and robustness in DAS-based activity recognition tasks. Referring to Fig. \ref{Fig.2}, it can be found that the result is at the same level as the SD case, which means that the two proposed solutions can almost completely alleviate the performance losses caused by environmental heterogeneity.

\section{Conclusions}

In this paper, we proposed and evaluated a lightweight, scalable approach for enabling active travel activity classification using AI-enpowered DAS systems. To address the challenge of heterogeneity caused by geographical environment and deployment across DAS nodes, we proposed an IoT solution based on FL and an initialization solution based on meta learning to improve the generalization capability of the system. The experimental results demonstrated that both solutions can improve the classification accuracy to the same level as local training, and the proposed solutions require fewer local training data and complexity, which is more suitable for rapid training and deployment of DAS systems.

\vspace{-0.2cm}
\footnotesize
\renewcommand{\baselinestretch}{2.0}
\bibliography{DAS_active_abb}
\bibliographystyle{IEEEtranN}

\end{document}